\documentclass[aps,preprint,preprintnumbers,nofootinbib,showpacs,superscriptaddress]{revtex4}
\usepackage{graphicx,color,amsmath}
\begin{document}

\def\lsim{ {\ \lower-1.2pt\vbox{\hbox{\rlap{$<$}\lower5pt\vbox{\hbox{$\sim$}
}}}\ } }
\def\gsim{ {\ \lower-1.2pt\vbox{\hbox{\rlap{$>$}\lower5pt\vbox{\hbox{$\sim$}
}}}\ } }

\font\el=cmbx10 scaled \magstep2{\obeylines\hfill September, 2008}
\vskip 1.5 cm
\title{Charmful Three-body Baryonic $B$ decays}
\author{Chun-Hung  Chen}
\affiliation{Department of Physics, National Tsing Hua University, Hsinchu, Taiwan 300, R.O.C.}
\author{Hai-Yang  Cheng}
\affiliation{Institute of Physics, Academia Sinica, Taipei, Taiwan 115, R.O.C.}
\author{C. Q. Geng}
\affiliation{Department of Physics, National Tsing Hua University, Hsinchu, Taiwan 300, R.O.C.}
\author{Y. K. Hsiao}
\affiliation{Institute of Physics, Academia Sinica, Taipei, Taiwan 115, R.O.C.}
\date{\today}
\begin{abstract}
We study the charmful three-body baryonic $B$ decays with $D^{(*)}$ or $J/\Psi$ in the final state. We explain the measured rates of $\bar
B^0\to n\bar p D^{*+}$, $\bar B^0\to p\bar p D^{(*)0}$, and $B^-\to
\Lambda \bar p J/\Psi$ and predict the branching fractions of $\bar B^0\to \Lambda\bar p D^{*+}$, $\bar B^0\to
\Sigma^0\bar p D^{*+}$, $B^-\to\Lambda \bar p D^{0}$, and
$B^-\to\Lambda \bar p D^{*0}$ to be of order
$(1.2,\;1.1,\;1.1,\;3.9)\times 10^{-5}$, respectively. They
are readily accessible to the $B$ factories.
\end{abstract}


\maketitle
\newpage
\section{introduction}
There are several salient features of the charmless three-body
baryonic $B$ decays $B\to {\bf B \bar B'} M$ with $\bf B$ being a
baryon. First,  a peak near the threshold area of the dibaryon
invariant mass spectrum has been observed in many baryonic $B$
decays \cite{Lambdappi_Belle, ppK(star)pi_Belle, AD_Belle,
Lambdapbargamma_Belle, LambdaLambdaK_Belle, ppK_Babar,
ADLambdapbarpi_Belle, ppKpi_Belle, pppi_Babar,ppKstar_Belle}. The so-called
threshold effect indicates that the $B$ meson is preferred to
decay into a baryon-antibaryon pair with low invariant mass
accompanied by a fast recoil meson. Second, none of the two-body
charmless baryonic $B$ decays has been observed so far and the
present limit on their branching ratios has been pushed to the level of
$10^{-7}$ \cite{BtoBB_Babar,BtoBB_Belle}. This means that
three-body final states usually have rates larger than their
two-body counterparts; that is, $\Gamma(B\to {\bf B \bar B'}
M)>\Gamma(B\to{\bf B\bar B'})$. This phenomenon can be understood
in terms of the threshold effect, namely, the invariant mass of
the dibaryon is preferred to be close to the threshold (for a
review, see \cite{HYreview}). The configuration of the  two-body
decay $B\to{\bf B\bar B'}$ is not favorable since its invariant
mass is $m_B$. In $B\to {\bf B \bar B'} M$ decays, the effective
mass of the baryon pair is reduced as the emitted meson can carry
away a large amount of the energies released in baryonic $B$ decays. Third, in the dibaryon rest frame the outgoing
meson tends to have a correlation with the baryon or the
antibaryon. Experimentally, the correlation can be studied by
measuring the Dalitz plot asymmetry or the angular distribution of
the baryon or the antibaryon in the dibaryon rest frame. For
example, the proton in the dibaryon rest frame of $B^-\to p\bar p
K^-$ is found to prefer to move collinearly with its meson partner
\cite{AD_Belle,ppK_Babar,ppKpi_Belle}, whereas
the proton and $\pi^-$ in $B^-\to p\bar p \pi^-$ move back to back most of the time
\cite{ppKpi_Belle}.

While various theoretical ideas
\cite{HouSoni,Chua_glueball,Rosner,HYpole,HYppD,HYchamBaryon,HYreview,FSI}
have been put forward to explain the low mass threshold
enhancement, this effect can be understood in terms of a simple
short-distance picture \cite{Suzuki}. One energetic $q\bar q$ pair
must be emitted back to back by a hard gluon in order to produce a
baryon and an antibaryon in the two-body decay. This hard gluon is
highly off mass shell and hence the two-body decay amplitude is
suppressed by order of $\alpha_s/q^2$. In the three-body baryonic
$B$ decays, a possible configuration is that the ${\bf B\bar B'}$
pair  is emitted collinearly against the meson. The quark and
antiquark pair emitted from a gluon is moving nearly in the same
direction. Since this gluon is close to its mass shell, the
corresponding configuration is not subject to the short-distance
suppression. This implies that the dibaryon pair tends to
have a small invariant mass.

While the above-mentioned short-distance picture predicts a
correct angular distribution pattern for the decays $B^-\to p\bar
p\pi^-$, $B^-\to \Lambda_c^+\bar p\pi^+$, and  $B^-\to \Lambda\bar
p\gamma$
\cite{ppKpi_Belle,LambdaCpbarpi_Belle,Lambdapbargamma_Belle} , it
fails to explain the angular correlation observed in
$B^-\to p\bar p K^-$ \cite{AD_Belle,ppKpi_Belle} and $B^-\to\Lambda\bar p\pi^-$
\cite{ADLambdapbarpi_Belle}. The intuitive argument
that the $K^-$ in the $p\bar p$ rest frame is expected to emerge
parallel to $\bar p$ is not borne out by experiment
\cite{HYreview}. Likewise, the naive argument that the pion has
no preference for its correlation with the $\Lambda$ or the $\bar
p$ in the decay $B^-\to\Lambda\bar p\pi^-$ is ruled out by the new
Belle experiment \cite{ADLambdapbarpi_Belle} in which a strong correlation between the $\Lambda$ and
the pion is seen. Therefore, although the short-distance picture
appears to describe the global features of the three-body baryonic
$B$ decays, it cannot explain the angular correlation enigma encountered in
the penguin-dominated processes $B^-\to p\bar p K^-$ and
$B^-\to\Lambda\bar p\pi^-$.

The study of the charmful baryonic $B$ decays $B\to {\bf B \bar
B'} M_c$ may help improve our understanding of the underlying
mechanism for the threshold enhancement and the angular distribution
in three-body decays. First, the aforementioned three features
also manifest themselves in $B\to {\bf B \bar B'} M_c$. An
enhancement at the low dibaryon mass has been seen, for example,
in the decay $B\to p\bar p D^{(*)}$
\cite{ppD(star)_Belle,ppD(star)_Babar}, while the Dalitz plot of
$B\to p\bar p D^{(*)}$ \cite{ppD(star)_Babar} with asymmetric
distributions signals a nonzero angular distribution asymmetry.
Therefore, $B\to {\bf B \bar B'} M_c$ is a mirror reflecting the
phenomena observed in $B\to {\bf B \bar B'} M$. Second, since
the decay rates of $B\to {\bf B \bar B'} M_c$ are 10-1000 times larger than that of $B\to {\bf B \bar B'} M$, in principle
this will render the detection of angular distributions and Dalitz
asymmetries in the latter easier. Theoretically, the relevant
topologic quark diagrams for charmful decays are simpler than the
charmless ones as the latter receive not only tree but also penguin
contributions. In this work, we shall therefore focus on $B\to {\bf
B \bar B'} M_c$ with $M_c=D^{(*)}$ or $J/\Psi$, and our first task
is to see if we can explain the branching fractions.

This paper is organized as follows. The experimental data for the decay rates of $B\to {\bf B \bar B'} M_c$ are collected in Sec. II. The formulism is set in Sec. III with special attention to various transition form factors. We then proceed to have a numerical analysis in
Sec. IV and discuss some physical results in Sec. V.  Section VI contains our conclusions.

\section{Experimental data}
In this section,  we summarize the measured branching ratios for the three-body charmful baryonic $B$ decays $B\to {\bf B \bar B'}M_c$:
\begin{eqnarray}\label{decay1}
{Br}(\bar B^0\to n\bar p D^{*+})&=&(14.5^{+3.4}_{-3.0}\pm 2.7)\times 10^{-4}\;\;\text{[CLEO]}\,,\nonumber\\
{Br}(\bar B^0\to p\bar p D^{0})
&=&(1.13\pm 0.06 \pm 0.08)\times 10^{-4}\;\;\text{[BaBar]}\,,\nonumber\\
&=&(1.18\pm 0.15 \pm 0.16)\times 10^{-4}\;\;\text{[Belle]}\,,\nonumber\\
{Br}(\bar B^0\to p\bar p D^{*0})
&=&(1.01\pm 0.10 \pm 0.09)\times 10^{-4}\;\;\text{[BaBar]}\,,\nonumber\\
&=&(1.20^{+0.33}_{-0.29}\pm 0.21)\times 10^{-4}\;\;\text{[Belle]}\,,\nonumber\\
{Br}(\bar B^0\to \Lambda\bar p D^{+}_s)
&=&(2.9\pm 0.7\pm 0.5\pm 0.4)\times 10^{-5}\;\;\text{[Belle]}\,,\nonumber\\
{Br}( B^-\to p\bar p D^{-})
&<&0.15\times 10^{-4}\;\;\text{(90\% C.L.) [Belle]}\,,\nonumber\\
{Br}(B^-\to p\bar p D^{*-})&<&0.15 \times 10^{-4}\;\;\text{(90\% C.L.) [Belle]}\,,
\end{eqnarray}
for $M_c=D^{(*)}$ \cite{Dstarpn_Cleo,ppD(star)_Belle,ppD(star)_Babar,DsLambdapbar_Belle},
and
\begin{eqnarray}  \label{decay2}
{Br}(B^-\to \Lambda\bar p J/\Psi)
&=&(11.6^{+7.4+4.2}_{-5.3-1.8})\times 10^{-6}\;\;\text{[BaBar]}\,,\nonumber\\
&=&(11.6\pm 2.8^{+1.8}_{-2.3})\times 10^{-6}\;\;\text{[Belle]}\,,\nonumber\\
{Br}(B^-\to \Sigma^0\bar p J/\Psi)
&<&1.1\times 10^{-5}\;\;\text{(90\% C.L.) [Belle]}\,,\nonumber\\
{Br}(\bar B^0\to  p\bar p J/\Psi)
&<&1.9\times 10^{-6}\;\;\text{(90\% C.L.) [BaBar]}\,,\nonumber\\
&<&8.3\times 10^{-7}\;\;\text{(90\% C.L.) [Belle]}\,,
\end{eqnarray}
for $M_c=J/\psi$ \cite{JLambdapbar_Belle,JLambdapbar_Babar}.
Among various measured decay modes, $\bar B^0\to
n\bar p D^{*+}$ was first observed by CLEO in 2001 \cite{Dstarpn_Cleo}. Note
that the decays $\bar B^0\to p\bar p D^{0}$ and $\bar B^0\to p\bar p D^{*0}$ have similar results in rates.

\section{Formalism}
The effective Hamiltonian responsible for charmful baryonic $B$ decays reads
\cite{Buchalla}
\begin{eqnarray}\label{H1}
{\cal H}_{eff}=\frac{G_F}{\sqrt 2}V_{cb}V_{q q'}^*(c_1 O_1+c_2O_2)+H.c.\,,
\end{eqnarray}
where  $V_{ij}$ are the CKM matrix
elements and $c_i$ the Wilson coefficients. The four-quark operators $O_i$ are defined by
\begin{eqnarray}\label{H2}
O_1=(\bar q'_\alpha q_\alpha)_{V-A}(\bar c_\beta
b_\beta)_{V-A}\;,\;\; O_2=(\bar q'_\alpha q_\beta)_{V-A}(\bar
c_\beta b_\alpha)_{V-A}\;,
\end{eqnarray}
where $(\bar q'_i q_j)_{V-A}$ denotes $\bar
q'_i\gamma_\mu(1-\gamma_5) q_j$ with $q'=d,s$, $q=u,c$, and $i,j$ the color indices.

To proceed, we shall adopt the generalized factorization approach \cite{Hamiltonian,ali} in which the vertex and penguin corrections to hadronic matrix elements of four-quark operators  are absorbed in the effective Wilson coefficients $c_i^{\rm eff}$ so that the $\mu$ dependence in transition matrix elements is smeared out:
\begin{eqnarray}
 \sum  c_i(\mu)\langle Q_i(\mu)\rangle \to\sum c_i^{\rm eff}\langle Q_i\rangle_{\rm
 VIA},
\end{eqnarray}
where the subscript VIA means that the hadronic matrix element is
evaluated under the vacuum insertion approximation.
This approach has been well applied to the study of three-body baryonic $B$ decays
\cite{HYppD,HYpole,HYradi2,HYchamBaryon,HYreview,HYchamBaryon2,ChuaHouTsai0,ChuaHouTsai,ChuaHouTsai2,ChuaHou,Tsai,Geng,CP_ppKstar,ppKstar_Geng,NF_Geng,BBgamma2_ChengYang,CPandT}
.
\begin{figure}[t!]
\centering
\includegraphics[width=4in]{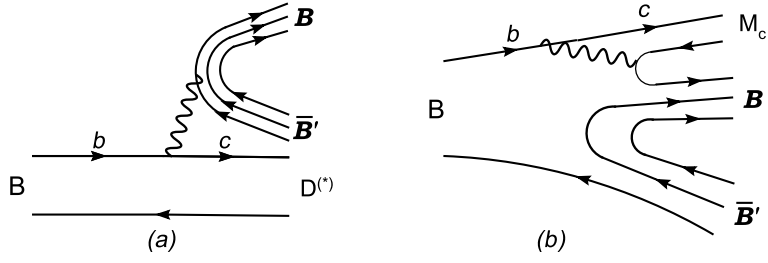}
\caption{Two types of the $B\to {\bf
B \bar B'} M_c$ decay process: (a) current type and (b) transition type.}\label{fig}
\end{figure}

Under the factorization approximation, the decay amplitudes can be classified into three different categories: the current-type (class-I), the transition-type (class-II), and the hybrid-type (class-III) amplitudes.
The class-I current-type amplitudes which proceed via a
color-allowed, external $W$-emission diagram as depicted in Fig.
\ref{fig}(a) are given by
\begin{eqnarray}\label{AC}
{\cal A_C}(B\to {\bf B\bar B'}D^{(*)})&=&\frac{G_F}{\sqrt 2}V_{cb}V_{uq'}^*a_1^{D^{(*)}} \langle{\bf B\bar B'}|(\bar q'u)_{V-A}|0\rangle\langle D^{(*)}|(\bar c b)_{V-A}|B\rangle\,,
\end{eqnarray}
with $a_1^{D^{(*)}}$ to be specified later.  By fixing $q=u$  to avoid the charm baryon production, all possible decay modes in this category are
\begin{eqnarray}\label{AC2}
&&\bar B^0\to\left\{ n\bar p     \;, \Sigma^-\bar\Lambda\;, \Lambda\overline{\Sigma^+}\;, \Sigma^-\overline{\Sigma^0}\;, \Sigma^0\overline{\Sigma^+}\;, \Xi^-\overline{\Xi^0}    \right\}D^{(*)+}\;\;\text{for $q'=d$}\;,\nonumber\\
&&\bar B^0\to\left\{\Lambda\bar p\;, \Sigma^0\bar p\;,\Sigma^-\bar n\;,\Xi^-\overline{\Sigma^0}\;,\Xi^-\bar\Lambda\;,          \Xi^0\overline{\Sigma^+}
\right\}D^{(*)+}\;\;\text{for $q'=s$}\;.
\end{eqnarray}
The class-II transition-type amplitude via
the color-suppressed internal $W$ emission diagram [Fig. \ref{fig}(b)] reads
\begin{eqnarray}\label{AT}
{\cal A_T}(B\to {\bf B\bar B'}M_c)&=&\frac{G_F}{\sqrt 2}V_{cb}V_{qq'}^*a_2^{M_c}\langle M_c|(\bar c q)_{V-A}|0\rangle\langle{\bf B\bar B'}|(\bar q' b)_{V-A}|B\rangle\,,
\end{eqnarray}
with $a_2^{M_c}$ to be given later.  The
possible decay modes in this class are
\begin{eqnarray}\label{AT2}
&&{B^-\to\left\{n\bar p\;,\Sigma^-\bar\Lambda\;,\Lambda\overline{\Sigma^+}\;,\Sigma^-\overline{\Sigma^0}\;, \Sigma^0\overline{\Sigma^+}\;,\Xi^-\overline{\Xi^0}\right\}J/\Psi}\;\;\text{for $q'=d$}\,,\nonumber\\
&&B^-\to\left\{\Lambda\bar p\;,\Sigma^0\bar p\;,\Sigma^-\bar n\;,\Xi^-\overline{\Sigma^0}\;, \Xi^-\bar\Lambda\;,\Xi^0\overline{\Sigma^+}\right\}J/\Psi\;\;\text{for $q'=s$}\,,
\end{eqnarray}
for the charged $B$ system, and
\begin{eqnarray}\label{AT3}
&{\bar B^0\to\left\{\begin{array}{l}
p\bar p\;,\;\;\Lambda\bar \Lambda\;,\;\;\Sigma^+\overline{\Sigma^+}\;,\Xi^-\overline{\Xi^-}\;,\Lambda\overline{\Sigma^0}\\
n\bar n\;,\Sigma^0\overline{\Sigma^0}\;,\Sigma^-\overline{\Sigma^-}\;,\Xi^0\overline{\Xi^0}\;,\;\;\Sigma^0\bar \Lambda
\end{array}\right\}D^{(*)0}(J/\Psi)\;\text{for $q'=d$}}\,,\nonumber\\
&\bar B^0\to\left\{\Lambda\bar n\;,\Sigma^0\bar n\;,\Sigma^+\bar p\;,\Xi^0\overline{\Sigma^0}\;,\Xi^0\bar \Lambda\;,\Xi^-\overline{\Sigma^-}\right\}D^{(*)0}(J/\Psi)\;\text{for $q'=s$}\,,
\end{eqnarray}
for the neutral $B$ system.
The class-III hybrid-type amplitudes which consist of both the current-type and the transition-type transitions are given by
\begin{eqnarray}\label{AH}
&&{\cal A_{H}}(B\to {\bf B\bar B'}D^{(*)})
=\frac{G_F}{\sqrt 2}V_{cb}V_{uq'}^*\bigg\{a_1^{D^{(*)}} \langle{\bf B\bar B'}|(\bar q'u)_{V-A}|0\rangle\langle D^{(*)}|(\bar c b)_{V-A}|B\rangle\nonumber\\
&&+a_2^{D^{(*)}}\langle D^{(*)}|(\bar c u)_{V-A}|0\rangle\langle{\bf B\bar B'}|(\bar q' b)_{V-A}|B\rangle\bigg\}\,.
\end{eqnarray}
The allowed modes are
\begin{eqnarray}\label{AH1}
&&B^-\to\left\{ n\bar p     \;, \Sigma^-\bar\Lambda\;, \Lambda\overline{\Sigma^+}\;, \Sigma^-\overline{\Sigma^0}\;, \Sigma^0\overline{\Sigma^+}\;, \Xi^-\overline{\Xi^0}    \right\}D^{(*)0}\;\;\text{for $q'=d$}\;,\nonumber\\
&&B^-\to\left\{\Lambda\bar p\;, \Sigma^0\bar p\;,      \Sigma^-\bar n\;,             \Xi^-\overline{\Sigma^0}\;,    \Xi^-\bar\Lambda\;,            \Xi^0\overline{\Sigma^+} \right\}D^{(*)0}\;\;\text{for $q'=s$}\;.
\end{eqnarray}

We now can have a qualitative understanding of the data
in (\ref{decay1}) and (\ref{decay2}). The nonobservation
of $B^-\to p\bar p D^{(*)-}$ is due to the fact that it proceeds via $b\to u\bar c d$ at the quark level. This leads
to a suppression of $|V_{ub} V_{cd}^*/V_{cb}V_{ud}^*|^2\simeq
10^{-4}$ compared to its neutral partner. Likewise, it is expected that $Br(\bar B^0\to p\bar p J/\Psi)=|V_{cd}/V_{cs}|^2\;{ Br}(B^-\to \Lambda\bar p
J/\Psi)\simeq 10^{-7}$, consistent with the experimental upper bound for this decay mode.
As for the comparable decay rates for $\bar B^0\to p\bar p
D^{*0}$ and $\bar B^0\to p\bar p D^{0}$,
it has to do with the baryonic form
factors, which we are going to elaborate on later.

\subsection{Decay constants and form factors}
To proceed, we need the information of the decay constants and form factors relevant for charmful baryonic $B$ decays. Decay constants are defined by
\begin{eqnarray}\label{dc}
\langle P_c|\bar q_1 \gamma_\mu \gamma_5 q_2|0\rangle&=&-if_{P_c} p_\mu\;,\nonumber\\
\langle V_c|\bar q_1 \gamma_\mu q_2|0\rangle&=&m_{V_c}f_{V_c}\varepsilon_\mu^*\;,
\end{eqnarray}
for the pseudoscalar $P_c$ and the vector meson $V_c$.
The $B$ to $D^{(*)}$ transition form factors are parametrized as \cite{BSW}
\begin{eqnarray}\label{ff1}
\langle D| \bar c \gamma^\mu b|B\rangle&=&\bigg[(p_B+p_D)^\mu-\frac{m^2_B-m^2_D}{t}q^\mu\bigg]F_1^{BD}(t)+\frac{m^2_B-m^2_D}{t}q^\mu F_0^{BD}(t)\,,\nonumber\\
\langle D^{*}|\bar c\gamma_\mu b|B\rangle&=&\epsilon_{\mu\nu\alpha\beta}
\varepsilon^{\ast\nu}p_B^{\alpha}p_{D^{*}}^{\beta}\frac{2V_1^{BD^*}(t)}{m_{B}+m_{D^{*}}}\;,\nonumber\\
\langle D^{*}|\bar c\gamma_\mu \gamma_5 b|B\rangle
&=&i\bigg[\varepsilon^\ast_\mu-\frac{\varepsilon^\ast\cdot q}{t}q_\mu\bigg](m_B+m_{D^{*}})A_1^{BD^*}(t)
 + i\frac{\varepsilon^\ast\cdot q}{t}q_\mu(2m_{D^{*}})A_0^{BD^*}(t)\nonumber\\
&-&i\bigg[(p_B+p_{D^{*}})_\mu-\frac{m^2_B-m^2_{D^{*}}}{t}q_\mu \bigg](\varepsilon^\ast\cdot q)\frac{A_2^{BD^*}(t)}{m_B+m_{D^{*}}}\;,
\end{eqnarray}
where $t\equiv q^2$ with $q=p_B-p_{D^{*}}=p_{\bf B}+p_{\bf\bar
B'}$.

For the dibaryon creation, we write
\begin{eqnarray}\label{timelikeF}
\langle {\bf B}{\bf\bar B'}|\bar q_1\gamma_\mu q_2|0\rangle
&=&\bar u(p_{\bf B})\bigg\{F_1(t)\gamma_\mu+\frac{F_2(t)}{m_{\bf B}+m_{\bf \bar B'}}i\sigma_{\mu\nu}q_\mu\bigg\}v(p_{\bf \bar B'})\;,\nonumber\\
&=& \bar u(p_{\bf B})\bigg\{[F_1(t)+F_2(t)]\gamma_\mu+\frac{F_2(t)}{m_{\bf B}+m_{\bf \bar B'}}(p_{\bf \bar B'}-p_{\bf B})_\mu\bigg\}v(p_{\bf \bar B'})\;,\nonumber\\
\langle {\bf B}{\bf\bar B'}|\bar q_1\gamma_\mu \gamma_5 q_2|0\rangle
&=&\bar u(p_{\bf B})\bigg\{g_A(t)\gamma_\mu+\frac{h_A(t)}{m_{\bf B}+m_{\bf \bar B'}}q_\mu\bigg\}\gamma_5 v(p_{\bf \bar B'})\,,
\end{eqnarray}
where $u$($v$) is the (anti-)baryon spinor, and $F_{1,2}$,
$g_A$, $h_A$ are timelike baryonic form factors. Note that there are two additional form factors in the form of $\bar uq_\mu v$ and $\bar u\sigma_{\mu\nu}q_\nu\gamma_5 v$. However, since we assume SU(3) flavor symmetry, we can neglect these two form factors as they vanish for conserved currents. The asymptotic behavior of form factors is governed by the pQCD
counting rules \cite{Brodsky1,Brodsky2,Brodsky3}. In the large $t$ limit,  the
momentum dependence of the form factors $F_1(t)$ and $g_A(t)$ behaves as $1/t^2$ as there are two hard gluon exchanges between the valence quarks. More precisely, in the $t\to\infty$ limit
\begin{eqnarray}\label{timelikeF2}
F_1(t)=\frac{C_{F_1}}{t^2}\bigg[\text{ln}\bigg(\frac{t}{\Lambda_0^2}\bigg)\bigg]^{-\gamma}\;, \qquad g_A(t)=\frac{C_{g_A}}{t^2}\bigg[\text{ln}\bigg(\frac{t}{\Lambda_0^2}\bigg)\bigg]^{-\gamma}\;,
\end{eqnarray}
where $\gamma=2+4/(3\beta)=2.148$ with $\beta$ being the QCD $\beta$ function and $\Lambda_0=0.3$ GeV.
In the asymptotic $t\to\infty$ limit, both $F_2(t)$ and $h_A(t)$ have an extra $1/t$ dependence relative to $F_1$ and $g_A$ owing to a mass insertion at the quark line \cite{Tsai,F2,F2b}. However,
the form factor $h_A$  is related to $g_A$ by the relation
\begin{eqnarray}\label{hA}
h_A=-\frac{(m_{\bf B}+m_{\bf \bar B'})^2}{t}g_A\;,
\end{eqnarray}
through the equation of motion. Hence, in ensuing numerical analysis we will keep $h_A(t)$ and neglect $F_2(t)$.
Under the SU(3) flavor and SU(2) spin symmetries \cite{Brodsky3}, the parameters $C_{F_1}$ and
$C_{g_A}$  appearing in $0\to {\bf B\bar B'}$ transitions are no longer
independent but are related to each other through the two reduced parameters $C_{||}$
and $C_{\overline{||}}$ (see Table \ref{timelikeF3}).

\begin{table}[t!]
\caption{The parameters $C_{F_1}$ and $C_{g_A}$ in Eq.
(\ref{timelikeF2}), expressed in terms of the combination of $C_{||}$
and $C_{\overline{||}}$, where the upper (lower) sign is for $C_{F_1}$ ($C_{g_A}$).} \label{timelikeF3}
\begin{tabular}{|c|c|}
\hline
$0\to$&$C_{F_1}$($C_{g_A}$)\\\hline
$n\bar p$                     &$\frac{4}{3}C_{||}\mp\frac{1}{3}C_{\overline{||}}$\\
$\Sigma^-\bar\Lambda$         &$\frac{1}{\sqrt 6}C_{||}\mp\frac{1}{\sqrt 6}C_{\overline{||}}$\\
$\Lambda\overline{\Sigma^+}$  &$\frac{1}{\sqrt 6}C_{||}\mp\frac{1}{\sqrt 6}C_{\overline{||}}$\\
$\Sigma^-\overline{\Sigma^0}$ &$\frac{5}{3\sqrt 2}C_{||}\pm\frac{1}{3\sqrt 2}C_{\overline{||}}$\\
$\Sigma^0\overline{\Sigma^+}$ &$\frac{-5}{3\sqrt 2}C_{||}\mp\frac{1}{3\sqrt 2}C_{\overline{||}}$\\
$\Xi^-\overline{\Xi^0}$       &$\frac{-1}{3}C_{||}\mp\frac{2}{3}C_{\overline{||}}$              \\
\hline
\end{tabular}
\begin{tabular}{|c|c|c|}
\hline
$0\to$&$C_{F_1}$($C_{g_A}$)\\\hline
$\Lambda\bar p$            &$-\sqrt{\frac{3}{2}}C_{||}$\\
$\Sigma^0\bar p$           &$\frac{-1}{3\sqrt 2}C_{||}\mp\frac{2}{3\sqrt 2}C_{\overline{||}}$\\
$\Sigma^-\bar n$           &$\frac{-1}{3}C_{||}\mp\frac{2}{3}C_{\overline{||}}$              \\
$\Xi^-\overline{\Sigma^0}$ &$\frac{4}{3\sqrt 2}C_{||}\mp\frac{1}{3\sqrt 2}C_{\overline{||}}$\\
$\Xi^-\bar\Lambda$         &$\frac{2}{\sqrt 6}C_{||}\pm\frac{1}{\sqrt 6}C_{\overline{||}}$  \\
$\Xi^0\overline{\Sigma^+}$&$\frac{4}{3}C_{||}\mp\frac{1}{3}C_{\overline{||}}$\\
\hline
\end{tabular}
\end{table}

For the three-body transition $B\to {\bf B\bar B'}$, its most general expression reads
\begin{eqnarray}\label{transitionF}
&&\langle {\bf B}{\bf\bar B'}|\bar q'\gamma_\mu b|B\rangle=\nonumber\\
&&i\bar u(p_{\bf B})[  g_1\gamma_{\mu}+g_2i\sigma_{\mu\nu}p^\nu +g_3p_{\mu} +g_4(p_{\bf\bar B'}+p_{\bf B})_\mu +g_5(p_{\bf\bar B'}-p_{\bf B})_\mu]\gamma_5v(p_{\bf \bar B'})\,,\nonumber\\
&&\langle {\bf B}{\bf\bar B'}|\bar q'\gamma_\mu\gamma_5 b|B\rangle=\nonumber\\
&&i\bar u(p_{\bf B})[ f_1\gamma_{\mu}+f_2i\sigma_{\mu\nu}p^\nu +f_3p_{\mu} +f_4(p_{\bf\bar B'}+p_{\bf B})_\mu +f_5(p_{\bf\bar B'}-p_{\bf B})_\mu]        v(p_{\bf \bar B'})\,.
\end{eqnarray}
with $p=p_B-p_{\bf B}-p_{\bf\bar B'}$. In principle, the form factors $f_i$ and $g_i$ depend on not only the invariant
mass $t$ of the dibaryon, but also the invariant mass (i.e. the variable $u$ or $s$) of one of the baryons and the emitted meson. Indeed, such a
momentum dependence  has been studied in the framework of
the pole model in which some intermediate pole contributions to the three-body transition of $B\to {\bf B\bar B'}$ are considered \cite{HYppD,HYpole,HYradi2,HYchamBaryon,HYreview}.
However, since the momentum dependence of the transition form factors on the variable $s$ or $u$ is poorly known, one approach which is often employed in the literature  is that  one first parametrizes them in the form of a power series of the dibaryon invariant mass $t$ and assumes that the dependence on other variables may be lumped into the coefficients of $1/t$ expansion
\cite{ChuaHouTsai2,Tsai,AngdisppK,Geng,CP_ppKstar,ppKstar_Geng}.
This is the approach we will adapt in this work.
 Since three gluons are needed to induce the $B\to{\bf B\bar B'}$ transition, two for producing the baryon pair and one for kicking the spectator quark in the $B$ meson,   pQCD
counting rules imply that to the leading order
\begin{eqnarray}\label{transitionF2}
f_i(t)=\frac{D_{f_i}}{t^3}\;, \qquad g_i(t)=\frac{D_{g_i}}{t^3}\;.
\end{eqnarray}

Just as the previous case for vacuum  to the dibaryon transition, under the SU(3) flavor and $SU(2)$ spin symmetries, the parameters $D_{g_1}$, $D_{f_1}$, $D_{g_i}$ and $D_{f_i}$ can be expressed in terms of the reduced parameters $D_{||}$, $D_{\overline{||}}$ and $D^i_{||}$
(see Table \ref{transitionF3}), and the derivation is presented in the Appendix. Interestingly, the decays $\bar B^0\to
\Sigma^+\overline{\Sigma^+}$ and $\Xi^0\overline{\Xi^0}$
are prohibited by SU(2) spin symmetry.

\begin{table}[t!]
\caption{ The parameters $D_{g_1}$, $D_{f_1}$, $D_{g_i}$, and $D_{f_i}$ in
$B\to{\bf B}{\bf \bar B'}$ transition expressed
in terms of the reduced parameters $D_{||}$, $D_{\overline{||}}$,
$D^i_{||}$ with $i=2, 3, 4, 5$,  where the upper (lower) sign is for $D_{g_1}$ ($D_{f_1}$). }\label{transitionF3}
\begin{tabular}{|l|c|c|}
\hline $B^-\to$&$D_{g_1}$($D_{f_1}$)&$D_{g_i},-D_{f_i}$\\\hline
$n\bar p$                     &$\frac{4}{3}D_{||}\pm\frac{1}{3}D_{\overline{||}}$&$\frac{5}{3}D^i_{||}$\\
$\Sigma^-\bar\Lambda$         &$\frac{1}{\sqrt 6}D_{||}\pm\frac{1}{\sqrt 6}D_{\overline{||}}$&$\sqrt{\frac{2}{3}}D^i_{||}$\\
$\Lambda\overline{\Sigma^+}$  &$\frac{1}{\sqrt 6}D_{||}\pm\frac{1}{\sqrt 6}D_{\overline{||}}$&$\sqrt{\frac{2}{3}}D^i_{||}$\\
$\Sigma^-\overline{\Sigma^0}$ &$\frac{5}{3\sqrt 2}D_{||}\mp\frac{1}{3\sqrt 2}D_{\overline{||}}$&$\frac{2\sqrt 2}{3}D^i_{||}$\\
$\Sigma^0\overline{\Sigma^+}$ &$-\frac{5}{3\sqrt 2}D_{||}\pm\frac{1}{3\sqrt 2}D_{\overline{||}}$&$-\frac{2\sqrt 2}{3}D^i_{||}$\\
$\Xi^-\overline{\Xi^0}$       &$-\frac{1}{3}D_{||}\pm\frac{2}{3}D_{\overline{||}}$&$\frac{1}{3}D^i_{||}$\\
\hline
\end{tabular}
\begin{tabular}{|l|c|c|}
\hline $B^-\to$&$D_{g_1}$($D_{f_1}$)&$D_{g_i},-D_{f_i}$\\\hline
$\Lambda\bar p$            &$-\sqrt{\frac{3}{2}}D_{||}$&$-\sqrt{\frac{3}{2}}D^i_{||}$\\
$\Sigma^0\bar p$           &$-\frac{1}{3\sqrt 2}D_{||}\pm\frac{2}{3\sqrt 2}D_{\overline{||}}$&$\frac{1}{3\sqrt 2}D^i_{||}$\\
$\Sigma^-\bar n$           &$-\frac{1}{3}D_{||}\pm\frac{2}{3}D_{\overline{||}}$&$\frac{1}{3}D^i_{||}$\\
$\Xi^-\overline{\Sigma^0}$ &$\frac{4}{3\sqrt 2}D_{||}\pm\frac{1}{3\sqrt 2}D_{\overline{||}}$&$\frac{5}{3\sqrt 2}D^i_{||}$\\
$\Xi^-\bar\Lambda$         &$\frac{2}{\sqrt 6}D_{||}\mp\frac{1}{\sqrt 6}D_{\overline{||}}$&$\frac{1}{\sqrt 6}D^i_{||}$\\
$\Xi^0\overline{\Sigma^+}$ &$\frac{4}{3}D_{||}\pm\frac{1}{3}D_{\overline{||}}$&$\frac{5}{3}D^i_{||}$\\
\hline
\end{tabular}
\begin{tabular}{|l|c|c|}
\hline
$\bar B^0\to$ &$D_{g_1}$($D_{f_1}$)&$D_{g_i},-D_{f_i}$\\
\hline
$^{p\bar p}_{\Xi^-\overline{\Xi^-}}\}$                    &$\frac{1}{3}D_{||}\mp\frac{2}{3}D_{\overline{||}}$     &$-\frac{1}{3}D^i_{||}$\\
$^{n\bar n}_{\Sigma^-\overline{\Sigma^-}}\}$              &$\frac{5}{3}D_{||}\mp\frac{1}{3}D_{\overline{||}}$     &$\frac{4}{3}D^i_{||}$ \\
$\Lambda\bar\Lambda$                                      &$\frac{1}{2}D_{||}\mp\frac{1}{2}D_{\overline{||}}$     &0                     \\
$\Sigma^0\overline{\Sigma^0}$                             &$\frac{5}{6}D_{||}\mp\frac{1}{6}D_{\overline{||}}$     &$\frac{2}{3}D^i_{||}$ \\
$^{\Sigma^+\overline{\Sigma^+}}_{\Xi^0\overline{\Xi^0}}\}$&0                                                    &0                     \\

$^{\Lambda\overline{\Sigma^0}}_{\Sigma^0\bar\Lambda}\}$   &$-\frac{1}{2\sqrt 3}D_{||}\mp\frac{1}{2\sqrt 3}D_{\overline{||}}$     &$-\frac{1}{\sqrt 3}D^i_{||}$\\
\hline
\end{tabular}
\begin{tabular}{|l|c|c|}
\hline
$\bar B^0\to$ &$D_{g_1}$($D_{f_1}$)&$D_{g_i},-D_{f_i}$\\
\hline
$\Lambda\bar n$           &$-\sqrt{\frac{3}{2}}D_{||}$&$-\sqrt{\frac{3}{2}}D^i_{||}$\\
$\Sigma^0\bar n$          &$\frac{1}{3\sqrt 2}D_{||}\mp\frac{2}{3\sqrt 2}D_{\overline{||}}$&$-\frac{1}{3\sqrt 2}D^i_{||}$\\
$\Sigma^+\bar p$          &$-\frac{1}{3}D_{||}\pm\frac{2}{3}D_{\overline{||}}$&$\frac{1}{3}D^i_{||}$\\
$\Xi^0\overline{\Sigma^0}$&$-\frac{4}{3\sqrt 2}D_{||}\mp\frac{1}{3\sqrt 2}D_{\overline{||}}$&$-\frac{5}{3\sqrt 2}D^i_{||}$\\
$\Xi^0\bar \Lambda$       &$\frac{2}{\sqrt 6}D_{||}\mp\frac{1}{\sqrt 6}D_{\overline{||}}$&$\frac{1}{\sqrt 6}D^i_{||}$\\
$\Xi^-\overline{\Sigma^-}$&$\frac{4}{3}D_{||}\pm\frac{1}{3}D_{\overline{||}}$&$\frac{5}{3}D^i_{||}$\\
\hline
\end{tabular}
\end{table}

Since many of the three-body baryonic $B$ decays show an enhancement at the low dibaryon
invariant mass squared $t$, low values of $t$ produce the dominant contributions to these decays. Recently, BaBar has measured the phase-space corrected dibaryon invariant mass  distributions for various $B\to D^{(*)}p\bar p$ decays \cite{ppD(star)_Babar}. This amounts to measuring the squared amplitude $|A|^2$ versus $t$. It turns out that the $t$ dependence of $|A|^2$ can be represented sufficiently by a single $1/t^n$ behavior. Therefore, it is justified to apply the pQCD counting rules valid at large $t$ to form factors at small values of $t$.


\section{Numerical Analysis}
We need to specify various input parameters for a numerical analysis.
For the CKM matrix elements, we use the Wolfenstein parameters
$A=0.807\pm0.018$, $\lambda=0.2265\pm0.0008$, $\bar \rho=0.141$ and $\bar
\eta=0.343$ \cite{CKMfitter}. For the decay constants, we use \cite{pdg,fDstar}
\begin{eqnarray}
&&(f_{D},\;f_{D^*},f_{J/\Psi})=(0.22,\;0.23,\;0.41)\;\text{GeV}\;.
\end{eqnarray}
Following \cite{BtoD}, the momentum dependence of $B \to D^{(*)}$ transition form factors in Eq. (\ref{ff1}) is parametrized as
 \begin{eqnarray} \label{eq:FFpara}
 F(q^2)=\,{F(0)\over 1-a(q^2/m_{B}^2)+b(q^2/m_{B}^2)^2}\,.
 \end{eqnarray}
We shall use the parameters $a$, $b$ and $F(0)$ obtained in \cite{BtoD} ( see Table
\ref{Formfactor}).

\begin{table}[h!]
\caption{Various $B\to D^{(*)}$ form factors taken from \cite{BtoD}.}\label{Formfactor}
\begin{tabular}{|c||cc|cccc|}
\hline $B\to D^{(*)}$
                 &$F_1$&$F_0$&$V_1$&$A_0$&$A_1$&$A_2$\\\hline
$F(0)$             &0.67 &0.67 &0.75 &0.64 &0.63 &0.61\\
$a$                &1.25 &0.65 &1.29 &1.30 &0.65 &1.14\\
$b$                &0.39 &0.00 &0.45 &0.31 &0.02 &0.52\\\hline
\end{tabular}
\end{table}

For the parameters $C_{||}$ and $C_{\overline{||}}$  in
Eqs. ({\ref{timelikeF}, \ref{timelikeF2}}) and Table
{\ref{timelikeF3}}, we use the data of  $e^+ e^-\to p\bar
p,\;n\bar n$ \cite{eetopp,eetonn} to determine their magnitudes
and the decay rate of $\bar B^0 \to n\bar p D^{*+}$ to fix their relative sign
\begin{eqnarray}\label{C1}
&&(C_{||},\,C_{\overline{||}})=(67.9\pm 1.4,\,-216.9\pm 23.5)\,{\rm GeV^{4}}\;,
\end{eqnarray}
with $\chi^2/d.o.f=1.6$.
As for the parameters $D_{||}$ and $D_{\overline{||}}$ in Eqs. (\ref{transitionF},
\ref{transitionF2}) and Table \ref{transitionF3},  we employ the observed rates of $\bar B^0 \to p\bar p D^{0}$,
$B^- \to p\bar p K^{*-}$, $\bar B^0 \to p\bar p K^{*0}$, and $B^-
\to p\bar p \pi^-$
\cite{ppK(star)pi_Belle,AD_Belle,ppKpi_Belle,pppi_Babar,ppD(star)_Belle,ppD(star)_Babar,ppKstar_Belle}
in conjunction  with the measured angular distribution of the last decay mode to obtain
\begin{eqnarray}
&&(D_{||},\;D_{\overline{||}})=(67.7\pm 16.3,\,-280.0\pm 35.9)\;{\rm GeV^5},\nonumber\\
&&(D_{||}^2,\,D_{||}^3,\,D_{||}^4,\,D_{||}^5)=\nonumber\\
&&(-187.3\pm 26.6,\,-840.1\pm 132.1,\,-10.1\pm 10.8,\,-157.0\pm 27.1)\;{\rm GeV^4}\;,
\end{eqnarray}
with $\chi^2/d.o.f=0.8$.
To calculate the decay rates, we use the relation
\begin{eqnarray}
d\Gamma=\frac{1}{(2\pi)^3}\frac{|\bar A|^2}{32M^3_B}dm^2_{\bf
B\bar B'}dm^2_{{\bf \bar B'}{M_c}}
\end{eqnarray}
with $m^2_{{\bf B\bar B'}}=(P_{\bf B}+P_{\bf \bar
B'})^2$, $m^2_{{\bf B'}{M_c}}=(P_{\bf \bar B'}+P_{M_c})^2$, and the baryon spins have been summed over in $|\bar A|^2$. The numerical
results for the branching ratios are summarized in Tables \ref{tab:r1}-\ref{r4}, where the theoretical errors come from the uncertainties in form factors.

\begin{table}[!]
\caption{Branching ratios for the current-type $\bar B^0\to{\bf B\bar
B'}D^{(*)+}$ decays. The results for $D^*$ production are shown in parentheses.}\label{tab:r1}
\begin{tabular}{|l|c|}
\hline ${\bf B\bar B'}$&$Br(\bar B^0\to{\bf B\bar
B'}D^{(*)+})\times 10^4$\\\hline
$n\bar p$                      &$5.2\,\;\pm 0.5\,\;\;\;\;(14.5\pm 1.4\;\,)$\\
$\Sigma^-\bar\Lambda$          &$0.3\,\;\pm 0.1\,\;\;\;\;( 1.3\;\,\pm 0.2\;\,)$ \\
$\Lambda\overline{\Sigma^+}$   &$0.3\,\;\pm 0.1\,\;\;\;\;(1.3\;\,\pm 0.2\;\,)$ \\
$\Sigma^-\overline{\Sigma^0}$  &$0.05\pm 0.01\;\;\;(0.23\pm 0.02)$ \\
$\Sigma^0\overline{\Sigma^+}$  &$0.05\pm 0.01\;\;\;(0.23\pm 0.02)$ \\
$\Xi^-\overline{\Xi^0}$        &$0.07\pm 0.02\;\;\;\;(0.4\;\pm0.1\;\;)$ \\
\hline
\end{tabular}
\begin{tabular}{|l|c|}
\hline ${\bf B\bar B'}$&$Br(\bar B^0\to{\bf B\bar B'}D^{(*)+})\times 10^6$\\\hline
$\Lambda\bar p$                &$3.4\pm 0.2\;\;\;(11.9\pm 0.5)$   \\
$\Sigma^0\bar p$               &$2.8\pm 0.7\;\;\;(10.5\pm 2.7)$   \\
$\Sigma^-\bar n$               &$5.5\pm 1.4\;\;\;(20.9\pm 5.4)$   \\
$\Xi^-\overline{\Sigma^0}$     &$0.6\pm 0.1\;\;\;(\;\;3.0\pm 0.3)$\\
$\Xi^-\bar\Lambda$             &$0.2\pm 0.1\;\;\;(\;\;1.3\pm 0.4)$\\
$\Xi^0\overline{\Sigma^+}$     &$1.2\pm 0.1\;\;\;(\;\;6.2\pm 0.6)$\\
\hline
\end{tabular}
\vskip 0.5 cm
\caption{Same as Table \ref{tab:r1} except for the transition-type  $\bar B^0\to{\bf B\bar
B'}D^{(*)0}$ decays.}\label{tab:r2}
\begin{tabular}{|l|c|}
\hline ${\bf B\bar B'}$&$Br(\bar B^0\to{\bf B\bar
B'}D^{(*)0})\times 10^4$\\\hline
$p\bar p$                                              &$1.1\;\;\pm 0.2\;\,$ \,\;\;\;$(1.0\;\;\pm 0.4\,\;)$ \\
$n\bar n$                                              &$3.9\;\;\pm 1.3\;\,$ \,\;\;\;$(5.8\;\;\pm 2.3\,\;)$          \\
$\Lambda\bar\Lambda$                                   &$0.02\pm 0.01$       \,\;\;\;$(0.03\pm 0.02)$         \\
$\Sigma^0\overline{\Sigma^0}$                          &$0.10\pm 0.03$       \,\;\;\;$(0.07\pm 0.03)$        \\
$\Sigma^-\overline{\Sigma^-}$                          &$0.4\;\,\pm 0.1\;\,$ \,\;\;\;$(0.3\;\;\pm 0.1\,\;)$         \\
$\Xi^-\overline{\Xi^-}$                                &$0.022\pm 0.004$ $(0.012\pm 0.005)$       \\
$\{^{\Lambda\overline{\Sigma^0}}_{\Sigma^0\bar\Lambda}$&$0.18\pm 0.05$    \,\,\;\;$(0.12\pm 0.04)$          \\
\hline
\end{tabular}
\begin{tabular}{|l|c|}
\hline ${\bf B\bar B'}$&$Br(\bar B^0\to{\bf B\bar
B'}D^{(*)0})\times 10^6$\\\hline
$\Lambda\bar n$           &$\;\;8.2\pm 2.7$ \,\;\;\;$(\;\;9.8\pm 3.7)$\\
$\Sigma^0\bar n$          &$\;\;0.7\pm 0.1$ \,\;\;\;$(\;\;0.6\pm 0.2)$\\
$\Sigma^+\bar p$          &$\;\;1.4\pm 0.3$ \,\;\;\;$(\;\;1.2\pm 0.4)$\\
$\Xi^0\overline{\Sigma^0}$&$\;\;1.2\pm 0.4$ \,\;\;\;$(\;\;0.7\pm 0.3)$\\
$\Xi^0\bar \Lambda$       &$\;\;0.09\pm 0.04$ $(\;\;0.10\pm 0.04)$ \\
$\Xi^-\overline{\Sigma^-}$&$\;\;2.2\pm 0.7$ \,\;\;\;$(\;\;1.4\pm 0.5)$ \\
&\\
\hline
\end{tabular}
\vskip 0.5 cm
\caption{Same as Table \ref{tab:r1} except for the hybrid-type $B^-\to{\bf B\bar B'}D^{(*)0}$ decays.}\label{tab:r3}
\begin{tabular}{|l|c|}
\hline ${\bf B\bar B'}$&$Br(B^-\to{\bf B\bar B'}D^{(*)0})\times
10^4$\\\hline
$n\bar p$                      &$13.0\pm 2.9$    \,$(46.6\pm 5.3)$\\
$\Sigma^-\bar\Lambda$          &$\;\;0.6\pm 0.1$ \;\;$(2.6\pm 0.3)$\\
$\Lambda\overline{\Sigma^+}$   &$\;\;0.6\pm 0.1$ \;\;$(2.6\pm 0.3)$\\
$\Sigma^-\overline{\Sigma^0}$  &$\;\;0.3\pm 0.1$ \;\;$(0.6\pm 0.1)$\\
$\Sigma^0\overline{\Sigma^+}$  &$\;\;0.3\pm 0.1$ \;\;$(0.6\pm 0.1)$\\
$\Xi^-\overline{\Xi^0}$        &$0.09\pm 0.01$ $(0.6\pm 0.1)$\\
\hline
\end{tabular}
\begin{tabular}{|l|c|}
\hline ${\bf B\bar B'}$&$Br(B^-\to{\bf B\bar B'}D^{(*)0})\times
10^6$\\\hline
$\Lambda\bar p$                &$11.4\pm 2.6$    \;\;$(32.3\pm 3.2)$\\
$\Sigma^0\bar p$               &$\;\;3.3\pm 0.6$ \;\;$(15.2\pm 2.4)$   \\
$\Sigma^-\bar n$               &$\;\;7.0\pm 1.2$ \;\;$(30.2\pm 4.7)$ \\
$\Xi^-\overline{\Sigma^0}$     &$\;\;1.9\pm 0.3$ \;\;$(\;\;6.2\pm 0.5)$ \\
$\Xi^-\bar\Lambda$             &$\;\;0.5\pm 0.1$ \;\;$(\;\;1.2\pm 0.2)$ \\
$\Xi^0\overline{\Sigma^+}$     &$\;\;3.7\pm 0.7$ \;\;$(\;12.7\pm 1.1)$ \\
\hline
\end{tabular}
\end{table}
\begin{table}[htb]
\caption{Branching ratios for the transition-type  $B^-\to{\bf B\bar
B'}J/\Psi$ decays.}\label{r4}
\begin{tabular}{|l|c|}
\hline ${\bf B\bar B'}$&$Br(B^-\to{\bf B\bar B'}J/\Psi)\times
10^6$\\\hline
$n\bar p$                      &$\,14.4\pm 3.2$\\
$\Sigma^-\bar\Lambda$          &$\;\;0.04\pm 0.02$\\
$\Lambda\overline{\Sigma^+}$   &$\;\;0.04\pm 0.02$\\
$\Sigma^-\overline{\Sigma^0}$  &$\;\;0.29\pm 0.09$\\
$\Sigma^0\overline{\Sigma^+}$  &$\;\;0.29\pm 0.09$\\
$\Xi^-\overline{\Xi^0}$        &$\;\;0.37\pm 0.08$\\
\hline
\end{tabular}
\begin{tabular}{|l|c|}
\hline ${\bf B\bar B'}$&$Br(B^-\to{\bf B\bar B'}J/\Psi)\times
10^6$\\\hline
$\Lambda\bar p$                &$11.6\pm \;2.9$\\
$\Sigma^0\bar p$               &$0.13\pm 0.02$\\
$\Sigma^-\bar n$               &$0.24\pm 0.05$\\
$\Xi^-\overline{\Sigma^0}$     &$22.9\pm \;5.8$\\
$\Xi^-\bar\Lambda$             &$0.52\pm 0.33$\\
$\Xi^0\overline{\Sigma^+}$     &$44.8\pm 10.8$\\
\hline
\end{tabular}
\vskip 0.5 cm
\caption{Branching ratios for the transition-type  $\bar B^0\to{\bf B\bar
B'}J/\Psi$ decays.}\label{r5}
\begin{tabular}{|l|c|}
\hline${\bf B\bar B'}$&$Br(\bar B^0\to{\bf B\bar B'}J/\Psi)\times
10^6$\\\hline
$p\bar p$                                              &$\;1.2\;\pm 0.2\;\;$\\
$n\bar n$                                              &$\;6.7\;\pm 1.7\;\;$\\
$\Lambda\bar\Lambda$                                   &$\;0.0010\;\pm \;\,0.0004$\\
$\Sigma^0\overline{\Sigma^0}$                          &$\;\;0.13\;\;\pm \;\;0.04\;\;$\\
$\Sigma^-\overline{\Sigma^-}$                          &$\;\;0.54\;\;\pm \;\;0.16\;\;$\\
$\Xi^-\overline{\Xi^-}$                                &$\;\;0.35\;\;\pm \;\;0.08\;\;$\\
$\{^{\Lambda\overline{\Sigma^0}}_{\Sigma^0\bar\Lambda}$&$\;\;0.09\;\;\pm \;\;0.02\;\;$\\
\hline
\end{tabular}
\begin{tabular}{|l|c|}
\hline${\bf B\bar B'}$&$Br(\bar B^0\to{\bf B\bar B'}J/\Psi)\times
10^6$\\\hline
$\Lambda\bar n$           &$10.4\pm 2.6$\\
$\Sigma^0\bar n$          &$\;\,0.11\pm 0.02$\\
$\Sigma^+\bar p$          &$\;\,0.24\pm 0.05$\\
$\Xi^0\overline{\Sigma^0}$&$20.6\pm 4.8$\\
$\Xi^0\bar \Lambda$       &$\;\,0.5\pm 0.3$\\
$\Xi^-\overline{\Sigma^-}$&\;$42.1\pm 10.4$\\
&\\
\hline
\end{tabular}
\end{table}

\section{Discussion}
In the generalized factorization approach, the coefficients $a_1$ and $a_2$ appearing in the decay amplitudes
$\cal A_{C}$, $\cal A_{T}$ and $\cal A_H$ are given by
\cite{Buchalla,Hamiltonian,ali}
\begin{eqnarray}
a_1=c_1^{\rm eff}+\frac{c_2^{\rm eff}}{3}+c_2^{\rm eff}\chi_1\,,\qquad a_2=c_2^{\rm eff}+\frac{c_1^{\rm eff}}{3}+c_1^{\rm eff}\chi_2\,,
\end{eqnarray}
with $(c_1^{\rm eff},\;c_2^{\rm eff})=(1.169,\;-0.367)$.  Since the nonfactorizable effects characterized by the $c_i^{\rm eff}\chi_j$ terms are not calculable, we will fit them to the data.
We see that $a_2$ is dominated by the nonfactorizable term due to the large cancelation between $c_2^{\rm eff}$ and $c_1^{\rm eff}/3$. Hence, nonfactorizable contributions to color-suppressed modes are sizable. Factorization works if the parameters $a_1$ and $a_2$ are universal; namely, they are channel by channel independent. Empirically, this is supported by the experimental measurements \cite{ai_HY}. Two-body mesonic $B$ decays suggest that
$a_1\sim {\cal O}(1)$ and $a_2\sim {\cal O}(0.2-0.3)$ \cite{ai_HY,ai_Neubert}.
It will become plausible if $a_i^{M_c}$ fitted from the
baryonic $B$ decays lie in the aforementioned ranges.
Indeed, the results given by
\begin{eqnarray}
a_1^{D^*}=1.23\pm 0.19\;,\;\;a_2^{D^*}=0.33\pm 0.04\;,\;\;a_2^{J/\Psi}=0.17\pm 0.03\;,
\end{eqnarray}
agree with the above expectation. Hence, we shall assume the validity of factorization in charmful baryonic $B$ decays.

There are several salient features we learn from Tables \ref{tab:r1}-\ref{tab:r3}: (i) ${\cal B}(\bar B^0\to {\bf B\bar B'}D^{*+})/{\cal B}(\bar B^0\to {\bf B\bar B'}D^{+})\gsim 3$
for class-I (current-type) decays; (ii) for class-II (transition-type) decays, ${\cal B}(\bar B^0\to {\bf B\bar B'}D^{*0})/{\cal B}(\bar B^0\to {\bf B\bar B'}D^{0})\gsim 1$ or $\lsim 1$, depending on the modes under consideration,
and (iii) ${\cal B}(B^-\to {\bf B\bar B'}D^{*0})/{\cal B}(B^-\to {\bf B\bar B'}D^{0})\gsim 3$ for hybrid-type decays except for ${\bf B\bar B'}=\Xi^-\overline{\Sigma^0}$, $\Xi^0\overline{\Sigma^+}$, and $\Xi^-\bar \Lambda$.

To get insight into the above-mentioned first feature, we take ${\bf B\bar B'}=n\bar p$ as an example. In the dibaryon rest frame we have
$\vec{p}_{\bf B}=-\vec{p}_{\bf\bar B'}$ and
$\vec{p}_B=\vec{p}_{D^{(*)}}$. As the branching fractions are dominated by the threshold effect, we can consider the dibaryon in the
nonrelativistic limit where $q_\mu=(p_{\bf B}+p_{\bf \bar B'})_\mu\approx
(2 m, 0, 0, 0)$ with $E_{\bf B}(E_{\bf \bar B'})\approx m$. Then
we find
\begin{eqnarray}
\langle {\bf B}{\bf\bar B'}|\bar q_1\gamma_\mu q_2|0\rangle &\approx& -2m\,(0,1,\pm i,0)F_1(4m^2)\,,\nonumber\\
\langle {\bf B}{\bf\bar B'}|\bar q_1\gamma_\mu \gamma_5 q_2|0\rangle &\approx& -2m(1,0,0,0)(g_A(4m^2)+h_A(4m^2))\,.
\end{eqnarray}
However, the vacuum to ${\bf B\bar B'}$ transition induced by the axial-vector current is suppressed as $h_A(4m^2)\approx -g_A(4m^2)$ followed from Eq. (\ref{hA}). From Eqs. (\ref{AC}) and (\ref{ff1}) we see that the decay amplitudes $\bar B^0\to n\bar p D^{*+}$ and $\bar B^0\to n\bar p D^{*0}$ are governed by the form factors $A_1^{BD^*}$ and $F_1^{BD}$, respectively. Thus the ratio
of these amplitudes is given by
\begin{eqnarray}
\frac{{\cal A}(\bar B^0\to n\bar p D^{*+})}{{\cal A}(\bar B^0\to n\bar p D^+)}
\approx \frac{m_B+m_{D^*}}{2m_D}\frac{E_{D^*}}{|\vec{p}_D|}\,,
\end{eqnarray}
where we have assumed $a_1^{D^*}=a_1^{D}$ and taken the approximation of $A_1^{BD^*}/F_1^{BD}\simeq 1$ inferred from Table \ref{Formfactor}.
Note that since numerically $(m_B+m_{D^*})^2/(4m_D^2)\sim 3$, we thus have ${\cal B}(\bar B^0\to n \bar p D^{*+})/{\cal B}(\bar B^0\to n \bar pD^{+})\sim 3$
as the ratio ${E_{D^*}}/{|\vec{p}_D|}$ is close to 1. For heavy dibaryons,
the ratio ${\cal B}(\bar B^0\to {\bf B\bar B'}D^{*+})/{\cal B}(\bar B^0\to {\bf B\bar B'}D^{+})$ will become larger, for example,
${\cal B}(\bar B^0\to \Xi^-\overline{\Xi^0} D^{*+})/{\cal B}(\bar B^0\to \Xi^-\overline{\Xi^0} D^{+})\sim[(m_B+m_{D^*})E_{D^*}/(2m_D |\vec{p}_D|)]^2=4.2$.
Among the current-type modes listed in Table \ref{tab:r1}, we notice that $\bar B^0\to \Lambda\bar p D^{*+}$ and $\bar B^0\to \Sigma^0\bar p D^{*+}$ have the largest rates
\begin{eqnarray}
{\cal B}(\bar B^0\to \Lambda\bar p D^{*+})&=&1.2\times 10^{-5}\,,\nonumber\\
{\cal B}(\bar B^0\to \Sigma^0\bar p D^{*+})&=&1.1\times 10^{-5}\,.
\end{eqnarray}
A search of them at $B$ factories is strongly encouraged. Note that the experimental studies of the angular distributions in the
charmful cases may help solve the angular correlation puzzle in $\bar B^0\to \Lambda\bar p \pi^+$ owing to
the same $0\to {\bf B\bar B'}$ transition form factors appearing in both cases.

Similar to the charmless cases of $B^-\to p\bar p K^{*-}$ and
$\bar B^0\to p\bar p K^{*0}$ \cite{CP_ppKstar,ppKstar_Geng}, the
decay rate of the class-II decay $\bar B^0\to p\bar p D^{*0}$ is dominated by the form factors $f_2$
and $g_2$ defined in  Eq. (\ref{transitionF}).
Contrary to the $D^*$ production, the decay $\bar B^0\to p\bar p D^0$ receives contributions from the form factors $f_3$ and $g_3$ accompanied by $m_D^2$. Considering the  dibaryon in the nonrelativistic limit, the ratio of the amplitudes reads
\begin{eqnarray} \label{eq:ratio}
\frac{{\cal A}(\bar B^0\to p\bar p D^{*0})}{{\cal A}(\bar B^0\to p\bar p D^0)}=
\frac{g_1 |\vec{p}_{D^*}|+f_1 E_{D^*}\text{sin}\theta_p+2g_2 m_{D^*}(\sqrt 2 |\vec{p}_{D^*}| \text{cos}\theta_p+m_{D^*} \text{sin}\theta_p)}
{g_1 E_{D}+f_1 |\vec{p}_{D}|\text{sin}\theta_p+g_3 m_{D}^2}\,.
\end{eqnarray}
This ratio is expected to be of the order of unity since
$g_1$ and $f_1$ terms in Eq. (\ref{eq:ratio}) are of the same order, and so are $g_2$ and $g_3$ terms.

Class-III decays will have rates larger than class-I and class-II ones if the relative phase between $a_1$ and $a_2$ is small so that their interference is constructive.
To proceed, we use $a_1^{D^{(*)}}$ and
$a_2^{D^{(*)}}$ extracted from $\bar B^0\to n\bar p D^{*+}$ and $\bar B^0\to
p\bar p D^{*0}$, respectively, and neglect their phases. The results are summarized in Table \ref{tab:r3}. For example,
\begin{eqnarray}\label{hydata}
{\cal B}(B^-\to\Lambda \bar p D^{0}) &=& 1.1\times 10^{-5}\,,\nonumber\\
{\cal B}(B^-\to\Lambda \bar p D^{*0})&=& 3.2\times 10^{-5}\,,
\end{eqnarray}
which are larger than ${\cal B}(\bar B^0\to\Lambda \bar p D^{+})$ and
${\cal B}(\bar B^0\to\Lambda \bar p D^{*+})$, respectively, by a factor of 3. It will be interesting to test this experimentally.

As seen in Table \ref{r4} and \ref{r5}, for the baryonic decays with a $J/\psi$ in the final state, our prediction ${\cal B}(B^-\to \Sigma^0\bar p J/\Psi)=1.3\times 10^{-7}$ is consistent with the Belle limit, $1.1\times 10^{-5}$, and ${\cal B}(\bar B^0\to p\bar p J/\Psi)=1.2\times 10^{-6}$ is in accordance with the BaBar limit but slightly higher than the upper bound set by Belle [see Eq. (\ref{decay2}) for the data].
We also find that the decays
$B^-\to ( \Xi^-\overline{\Sigma^0},\; \Xi^-\Lambda,\;
\Xi^0\overline{\Sigma^+,}\; \Xi^0\overline{\Sigma^0},\;
\Xi^-\overline{\Sigma^-} )J/\Psi$ have branching fractions of order $10^{-5}$.

\begin{figure}[t!]
\centering
\includegraphics[width=2.0in]{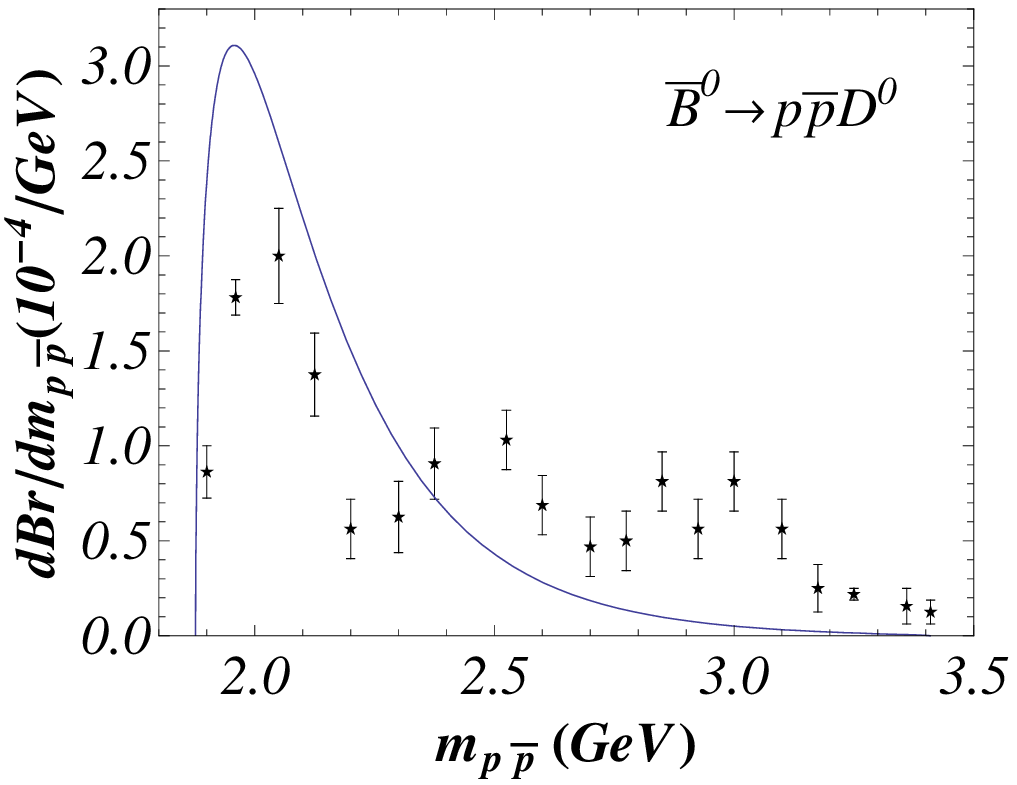}
\includegraphics[width=2.0in]{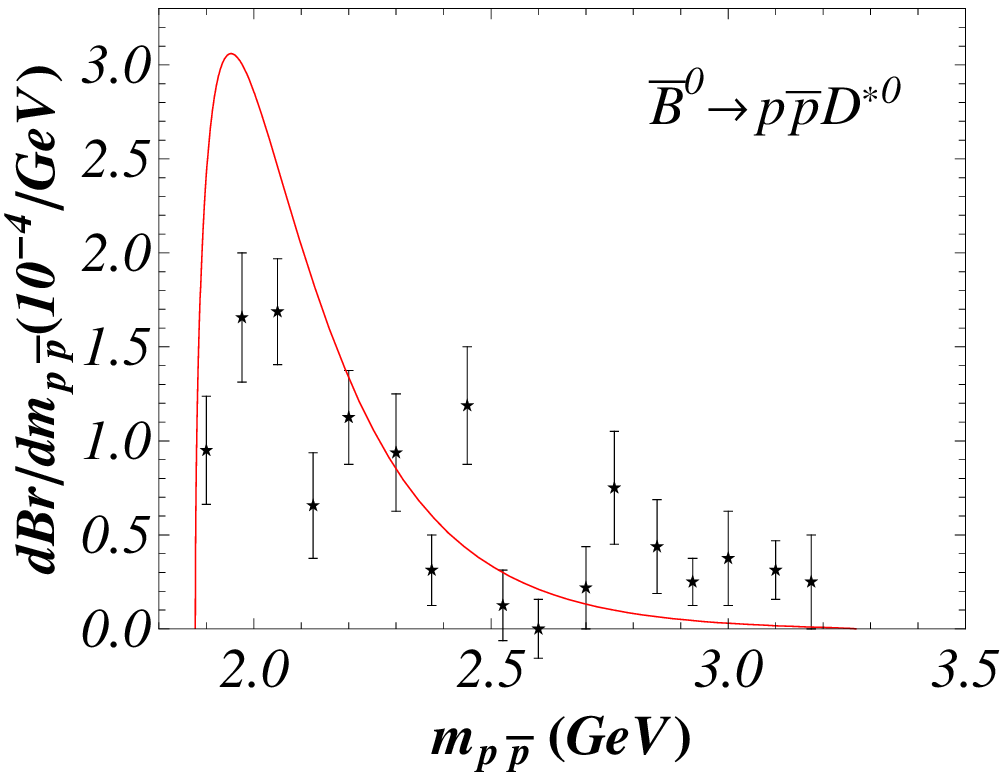}
\includegraphics[width=2.03in]{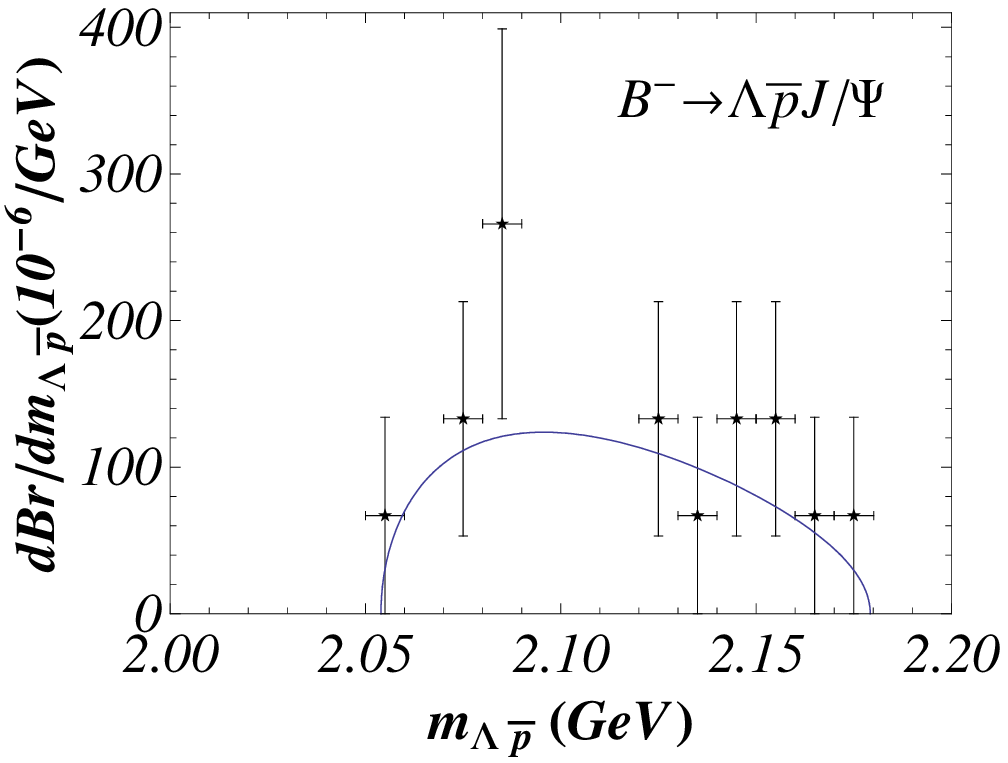}
\caption{Dibaryon invariant mass distributions for $\bar B^0\to
p\bar p D^{0}$, $\bar B^0\to
p\bar p D^{*0}$ and $B^-\to \Lambda \bar p J/\Psi$, respectively.
Experimental data are taken from \cite{ppD(star)_Babar,JLambdapbar_Belle}.}\label{fig_threshold}
\end{figure}

As for the threshold peaking effect,
while it manifests in the decay $\bar B^0\to p\bar p D^{(*)0}$ the data clearly do not show the threshold behavior in $B^-\to \Lambda\bar p J/\Psi$ (see Fig. \ref{fig_threshold}).
This can be understood as follows.
In the latter decay, the invariant mass $m_{\Lambda\bar p}$ ranges from 2.05 to 2.18 GeV, which is very narrow compared to the $m_{p\bar p}$ range in the
$\bar B^0\to p\bar p D^{(*)0}$ decay. Consequently, the invariant mass distribution of $d\Gamma/dm_{\Lambda\bar p}$
is governed by the shape of the phase space due to the relative flat $1/t^3$ dependence within the small allowed $m_{\Lambda\bar p}$ region.

Finally we notice that since the transition-type decays share the same $B\to
{\bf B\bar B'}$ transition form factors with $B^-\to p\bar p
\pi^-$, $B^-\to p\bar p K^-$ and $B^-\to \Lambda\bar p \gamma$,
measurements for distributions/Daltiz plot
asymmetries are useful for solving the puzzle of the angular
distributions in the charmless baryonic $B$ decays.
For example, a recent Dalitz plot analysis of $\bar B^0\to p\bar p D^{(*)0}$ by  BaBar \cite{ppD(star)_Babar} shows a possible indication that
\begin{eqnarray} \label{AppD}
A_\theta(\bar B^0\to p\bar p D^{0})>A_\theta(\bar B^0\to p\bar p D^{*0})>0\,,
\end{eqnarray}
where the angular asymmetry $A_\theta$ is defined by
\cite{AngdisppK}
\begin{eqnarray}\label{AFB}
A_{\theta}&\equiv&\frac{\int^{+1}_0\frac{d\Gamma}{d\cos\theta}d\cos\theta
-\int^0_{-1}\frac{d\Gamma}{d\cos\theta}
d\cos\theta}{\int^{+1}_0\frac{d\Gamma}{d\cos\theta}
d\cos\theta+\int^0_{-1}\frac{d\Gamma}{d\cos\theta} d\cos\theta}\;,
\end{eqnarray}
which is equal to $(N_+-N_-)/(N_++N_-)$, where $N_\pm$ are the
events with $\cos\theta>0$ and $\cos\theta<0$, respectively.
The above relation (\ref{AppD})
can be explained in our model: in the previous study in \cite{AngdisppK,ppKstar_Geng} we have predicted that  $|A_\theta(B^-\to p\bar p K^-)|>|A_\theta(B^-\to p\bar p K^{*-})|$ and $|A_\theta(B^-\to p\bar p \pi^-)|>|A_\theta(B^-\to p\bar p \rho^-)|$. In our model, the charmful baryonic $B$ decays  have similar features of the angular distribution as the charmless cases.
Therefore, we believe that our model together with the choice of the form factors will help understand the angular correlations observed in these decays.

\section{Conclusions}
Within the framework of  the generalized factorization approach, we have studied  three different types of charmful three-body baryonic $B$ decays with
$D^{(*)}$ or $J/\Psi$ in the final state. We found that the $B\to {\bf
B\bar B'}$ form factors extracted mainly from the data of charmless baryonic $B$ decays
can be used to explain the measured rates of $\bar B^0\to
p\bar p D^{(*)0}$ and $B^-\to \Lambda \bar p J/\Psi$.
The branching fractions of $\bar B^0\to \Lambda\bar p
D^{*+}$, $\bar B^0\to \Sigma^0\bar p D^{*+}$, $B^-\to\Lambda \bar
p D^{0}$, and $B^-\to\Lambda \bar p D^{*0}$ are predicted to be
$(1.2,\;1.1,\;1.1,\;3.9)\times 10^{-5}$, respectively. They
are readily accessible to $B$ factories.

\section*{Acknowledgments}
The authors would like thank Min-Zu Wang for useful discussions. This work is
supported in part by the National Science Council of R.O.C. under
Grants No. NSC96-2112-M-001-003 and  No. NSC-95-2112-M-007-059-MY3.

\section*{Appendix: Relations for $B\to {\bf B\bar B'}$ transition form factors}
Form factors of $\langle {\bf B}{\bf\bar B'}|\bar q\gamma_\mu
b|B\rangle$ and $\langle {\bf B}{\bf\bar B'}|\bar
q\gamma_\mu\gamma_5 b|B\rangle$ can be related to each other by the SU(3)
flavor and SU(2) spin symmetries in the large $t$ limit ($t\to \infty$).
As the currents $V_\mu(A_\mu)=\bar q\gamma_\mu(\gamma_5) b$ are used to
create the dibaryon and annihilate the $B$ meson with $|B\rangle\sim |\bar b\gamma_5 q'|0\rangle$, we have
\begin{eqnarray}
\langle {\bf B\bar B'}|V_\mu-A_\mu|B \rangle
&=&\langle {\bf B\bar B'}|2\bar q_L\gamma_\mu b_L(\bar b_L q'_R-\bar b_R q'_L)|0 \rangle\nonumber\\
&=&\langle {\bf B\bar B'}|2\bar q_L\gamma_\mu (\not{\!p}_b+m_b) q'_R-2\bar q_L\gamma_\mu(\not{\!p}_b+m_b) q'_L|0 \rangle\nonumber\\
&=&\langle {\bf B\bar B'}|-2m_b\bar q_L\gamma_\mu q'_L+2\bar q_L\gamma_\mu\not{\!p}_b q'_R|0 \rangle\nonumber\\
&=&\langle {\bf B\bar B'}|J_\mu|0 \rangle+\langle {\bf B\bar B'}|J'_\mu|0 \rangle\,,
\end{eqnarray}
where $J_\mu=-2m_b\bar q_L\gamma_\mu {q'_L}$ and
$J'_\mu=2\bar q_L\gamma_\mu\not\!\!p_b {q'_R}$. Note that
$J_\mu$ ($J'_\mu$) is the source of the chiral-even
(chiral-odd) current. In terms of the crossing symmetry, the
antibaryon in the final state can be transformed as the baryon in
the initial state.  Then we can follow Refs. \cite{Brodsky3,ChuaHou,Tsai} to parameterize the
${\bf B'\to B}$ transition form factors as
\begin{eqnarray}\label{J1J2}
\langle {\bf B}|J_\mu|{\bf B'} \rangle&=&2i m_b\bar u\gamma_\mu\bigg[\frac{1+\gamma_5}{2}D^+ +\frac{1-\gamma_5}{2}D^-\bigg]u\,,\nonumber\\
\langle {\bf B}|J'_\mu|{\bf B'} \rangle&=&2i\bar u\gamma_\mu\not{\!p}_b\bigg[\frac{1+\gamma_5}{2}D'^+ +\frac{1-\gamma_5}{2}D'^-\bigg]u\,.
\end{eqnarray}
The above chiral-even and chiral-odd form factors
$D^\pm$ and $D'^\pm$ are given by
\begin{eqnarray}
D^\pm&=&e_{||}^{\pm} D_{||}+e_{\overline{||}}^{\pm} D_{\overline{||}}\,,\nonumber\\
D'^\pm&=&e_{||}^{\pm} D'_{||}+e_{\overline{||}}^{\pm} D'_{\overline{||}}\,,
\end{eqnarray}
where $D_{{||}(\overline{||})}$, $D'_{{||}(\overline{||})}$
are the amplitudes in which the currents interact with a valence
quark with helicity parallel (antiparallel) to the helicity of
the baryon. The parameters $e_{||}^{\pm}$ and
$e_{\overline{||}}^\pm$ are defined by
\begin{eqnarray}
e_{||}^\pm           &=&\sum_{i<j}\langle {\bf B},h|Q(i)+Q(j)|{\bf B'},{h'}\rangle\,,\nonumber\\
e_{\overline{||}}^\pm&=&\sum_k\langle {\bf B},h|Q(k)|{\bf B'},{h'}\rangle\,,
\end{eqnarray}
where the charge $Q=J_0$ with $i,j,k=1,2,3$ numbering the valence
quarks in the baryon to be acted. Therefore,
$e_{||(\overline{||})}^\pm$ is the sum of the electroweak charges
carried by valence quarks in the baryon with helicities parallel
(antiparallel) to the baryon's helicity. Besides, the superscript
$+(-)$ denotes $h'=\uparrow(\downarrow)$, the helicities of the
initial baryon.

Take $\bar B^0\to p\bar p$ as an example. With the
helicity state of the proton given by
\begin{eqnarray}
|p,{\uparrow}\rangle&=&|p,\uparrow\downarrow\uparrow\rangle+|p,\uparrow\uparrow \downarrow\rangle+|p,\downarrow\uparrow\uparrow\rangle\nonumber\\
&=&\frac{1}{\sqrt 3}\left\{
\begin{array}{c}
\;\;\frac{1}{\sqrt 6}[2u_\uparrow d_\downarrow u_\uparrow-u_\uparrow u_\downarrow d_\uparrow-d_\uparrow u_\downarrow u_\uparrow]\\
+\frac{1}{\sqrt 6}[2u_\uparrow u_\uparrow d_\downarrow-u_\uparrow d_\uparrow u_\downarrow-d_\uparrow u_\uparrow u_\downarrow]\\
+\frac{1}{\sqrt 6}[2d_\downarrow u_\uparrow u_\uparrow-u_\downarrow d_\uparrow u_\uparrow-u_\downarrow u_\uparrow d_\uparrow]\\
\end{array}
\right\}\,,\nonumber\\
|p,\downarrow\rangle&=&(|p,\uparrow\rangle \;\text{with $\uparrow\leftrightarrow \downarrow$})\,,
\end{eqnarray}
we obtain
\begin{eqnarray}\label{QJ1}
e_{||}^-           &=&\sum_{i<j}\langle p,\downarrow|Q(i)+Q(j)|p,\downarrow\rangle=\frac{1}{3}\,,\nonumber\\
e_{||}^+           &=&\sum_{i<j}\langle p,\uparrow|Q(i)+Q(j)|p,\uparrow\rangle=0\,,\nonumber\\
e_{\overline{||}}^-&=&\sum_{k}\langle p,\downarrow|Q(k)|p,\downarrow\rangle=0\,,\nonumber\\
e_{\overline{||}}^+&=&\sum_{k}\langle p,\uparrow|Q(k)|p,\uparrow\rangle=\frac{2}{3}\,,
\end{eqnarray}
for the chiral-even $J_\mu$, and
\begin{eqnarray}\label{QJ2}
e_{||}^-           &=&\sum_{i<j}\langle p,\downarrow|Q(i)+Q(j)|p,\downarrow\rangle=0\,,\nonumber\\
e_{||}^+           &=&\sum_{i<j}\langle p,\uparrow|Q(i)+Q(j)|p,\uparrow\rangle=\frac{-1}{3}\,,\nonumber\\
e_{\overline{||}}^-&=&\sum_{k}\langle p,\downarrow|Q(k)|p,\downarrow\rangle=0\,,\nonumber\\
e_{\overline{||}}^+&=&\sum_{k}\langle p,\uparrow|Q(k)|p,\uparrow\rangle=0\,,
\end{eqnarray}
for the chiral-odd $J'_\mu$.
When the sums of the electroweak charges
$e_{||(\overline{||})}^\pm$ derived in Eqs. (\ref{QJ1},
\ref{QJ2}) are put back into Eq. (\ref{J1J2}), $\langle {\bf
B}|J_\mu+J'_\mu|{\bf B'} \rangle$ is given as
\begin{eqnarray}
\langle {\bf B}|J_\mu+J'_\mu|{\bf B'} \rangle
&=&i\bar u\bigg\{m_b[\frac{1}{3}D_{||}-\frac{2}{3}D_{\overline{||}}]\gamma_\mu+[-\frac{1}{3}D'_{||}]
\gamma_\mu\not{\!\!p}_b\bigg\}\gamma_5 u\,,\nonumber\\
&+&i\bar u\bigg\{m_b[\frac{1}{3}D_{||}+\frac{2}{3}D_{\overline{||}}]\gamma_\mu +[\frac{1}{3}D'_{||}]\gamma_\mu\not{\!\!p}_b\bigg\}u\,.
\end{eqnarray}
This gives the form factor relations for $\langle {\bf B}{\bf\bar B'}|\bar q\gamma_\mu
b|B\rangle$ and $\langle {\bf B}{\bf\bar B'}|\bar
q\gamma_\mu\gamma_5 b|B\rangle$, which is given by
\begin{eqnarray}
g_1&=&\frac{1}{3}D_{||}-\frac{2}{3}D_{\overline{||}}\,,\;f_1=\frac{1}{3}D_{||}+\frac{2}{3}D_{\overline{||}}\,,\nonumber\\
g_i&=&-f_i=-\frac{1}{3}D^i_{||}\,, \qquad\quad (i=2,\cdots,5).
\end{eqnarray}
By the same token, we find the relation of $g_i=-f_i$ holds in all $B\to {\bf B\bar B'}$ transition form factors.

\end{document}